\documentstyle[12pt]{article}
\begin{document}
\begin{titlepage}
\title{
{\bf Period of K system generator of pseudorandom numbers}
\footnote{Crete University reprint CRETE.TH/12/95; Hep-lat 9601003}
}
{\bf
\author{N.Z.Akopov\\
Yerevan Physics Institute,375036 Yerevan,Armenia\vspace{1cm}\\
G.G.Athanasiu\\
Physics Department,University of Crete,71409 Iraklion,Greece\vspace{1cm}\\
E.G.Floratos\\
National Research Center,$15310$ A. Paraskevi, Greece;\\
Physics Department,University of Crete,71409 Iraklion,Greece\vspace{1cm}\\
G.K.Savvidy\\
Physics Department,University of Crete,71409 Iraklion,Greece;\\
Research Foundation of Thrace, Interdisciplinary Center for Complexity
}
}
\date{}
\maketitle
\begin{abstract}
\noindent

We analyze the structure of the periodic trajectories of the matrix 
generator of pseudorandom numbers which has been earlier proposed in 
\cite{savvidy1,akopov1}. The structure of the periodic trajectories
becomes more transparent when the rational sublattice coincides with 
the Galois field $GF[p]$~~~\cite{athanas,athanas1}. We are able to 
compute the period of the trajectories as a function of $p$ and the 
dimension of the matrix $d$.
\end{abstract} 
\thispagestyle{empty}
\end{titlepage}
\pagestyle{empty}

\vspace{.5cm}

{\bf 1. }In the articles \cite{savvidy1,akopov1} 
the authors suggested the matrix generator of pseudorandom 
numbers based on Kolmogorov-Anosov K systems 
\cite{kolmogorov,anosov,rohlin,sinai}. 
These systems are the most stochastic 
dynamical systems, with nonzero Kolmogorov entropy 
\cite{kolmogorov,anosov,fomin,savvidy2,savvidy3,abramyan}. 
The trajectories of the system are exponentially unstable and 
uniformly fill the phase space. The coordinates of these 
trajectories represent the desired sequence of pseudorandom numbers
\cite{savvidy1,akopov1}.

In the given case it is assumed that the unit $d$-dimensional
torus $\Pi^{d}$ plays the role of the phase space, therefore the 
coordinates of the trajectory are uniformly distributed over the 
unit hypercube and can be used for Monte-Karo simulations 
\cite{savvidy1,akopov1}.

The properties of this new class of matrix
generators were investigated with different criterion 
including discrepancy $D_{N}$  in various dimensions. In all 
cases it shows good statistical properties \cite{akopov1}.
Matrix generators based on different ideas are proposed in 
\cite{niederreiter,grothe}. We refer to the book of Niederreiter
\cite{nied} and to the survey article \cite{niede} for recent 
references.

The aim of this article is to estimate the period of the 
trajectories which are used to produce pseudorandom numbers. 
Only periodic trajectories of the system can be simulated on 
a computer, because trajectories on a computer are always on 
a finite rational sublattice $Z_{p}$ of the phase space $\Pi^{d}$ 
\cite{dyson,vivaldy,pregrad}. 

The  essence of the approach is to consider the system on rational 
sublattice $Z_{p}$ of a unit $d$-dimensional torus and particularly 
on sublattices with $prime$ basis $p$ \cite{athanas,vivaldy}. 
These sublattices are equivalent to Galois fields 
$GF[p]$ and all four elementary arithmetical operations can be 
carried out unrestrictevely \cite{apostol,lidl,bastida}.
This approach demonstrates that in order to have a 
large period on a 
sublattice $GF[p]$ one should have matrices $A$ which 
have the eigenvalues in high extensions $GF[\sqrt[d]{p}]$ of
the field and at the same time they should have integer entries.

On every Galois field exists a primitive element $g$ which has 
the period $p-1$,~that is $g^{p-1}=1$,~ and every element of 
$GF[p]$ can be 
represented as a power of $g$ \cite{apostol,lidl,bastida}. 
We shall show that if the eigenvalues 
of the matrix generator are proportional to these primitive elements, 
then the period of the trajectories is equal to

$$\tau_{p}(A_{2}) = p-1.$$
The upper bound for the period of the cat map (15) 
found by Dyson and Falk \cite{dyson} 
on arbitrary rational sublattice $Z_{q}$ is equal to 

$$\tau \leq 3q $$
and also linearly depends on $q$.

The period increases when we consider the sublattices 
$Z_{p^{n}}$ of the basis 
$p^{n}$,~~where~~~$n>2$ instead of $p$. In that case the period of the 
same matrix generator is 

$$\tau_{p^{n}}(A_{2})= (p-1)p^{n-1}.$$
Then increasing the dimension of the matrix generator by 
$\{2d\times 2d\}$ matrices we will get the period equal to 

$$\tau_{p}(A_{2d}) = d(p-1)$$
and correspondingly on $Z_{p^{n}}$ sublattice 

$$\tau_{p^{n}}(A_{2d})= d(p-1)p^{n-1}.$$
Using quadratic extension of the Galois field $GF[\sqrt {p}]$ 
\cite{apostol,lidl,bastida} we
have found a systematic way to construct the matrix generators with period 

$$\tau_{p}= p^{2}-1, ~~~~~~and~~~~~~~~\tau_{p^{n}} =(p^{2}-1)p^{n-1}$$
and finally on the high extended Galois field $GF[\sqrt[d]{p}]$ 
the period is equal to 

$$\tau_{p}= p^{d}-1, ~~~~~~and~~~~~~~~\tau_{p^{n}} =(p^{d}-1)p^{n-1}.$$
{\it The last result shows that in practical simulations the period is 
very large and is of order $p^{d+n-1}$ where $d$- is the dimension of the 
matrix generator and $p^{n}$- is the basis of the sublattice $Z_{p^{n}}$.}

We suggest specific matrices with this properties and with
almost zero entries, see Section 17.  This matrices 
have the largest period 
and can be easily used for practical simulations. 

\vspace{.5cm}

{\bf 2. }Let us pass to the details of the algorithm. 
The  matrix  generator is defined as \cite{savvidy1,akopov1},

$$X^{(n+1)} =A \cdot  X^{(n)},~~~~~~~~~~~~ (mod~1),\eqno(1)$$
where $A$ is $d \times d$ dimensional matrix with integer 
matrix elements $a_{i,j}$ and determinant equal to one

$$Det~A = 1 ,\eqno(2)$$
and $X^{(0)}=(X^{(0)}_{1},...,X^{(0)}_{d})$  is an  initial vector. 
The last condition provides phase volume conservation.
The automorphism (1) form the $K$ system of Anosov if and only if 
all eigenvalues of the matrix $A$ are in modulus different 
from unity \cite{anosov,rohlin,sinai}

$$\vert \lambda_{i} \vert \neq 1,~~~~~~~i=1,...,d \eqno(3)$$ 
The $\it trajectory$ of the $K$ system (1) 

$$X_{0},X_{1},X_{2}....$$ 
represents the desired sequence of the pseudorandom numbers  
\cite{savvidy1}. 

This approach allows a 
large freedom in choosing the matrices $A$ for the K system 
generators and the initial vectors \cite{savvidy1}.
Specific choices suggested in \cite{savvidy1,akopov1,nersesian} are

$$ A_{d} = \left( \begin{array}{c}

        2,3,4,.......,d~~,1 \\
        1,2,3,.....,d-1,1 \\
        1,1,2,.....,d-2,1 \\
        ................. \\
        ................. \\
        1,1,1,...,2,3,4,1 \\
        1,1,1,...,1,2,2,1 \\
        1,1,1,...,1,1,2,1 \\
        1,1,1,...,1,1,1,1 
\end{array} \right) , A_{d}= \left( \begin{array}{c}
         0,~~1~,~~0~,.....,~~0 \\
         0,~~0~,~~1~,.....,~~0   \\
         .............   \\
         ............. \\
         0,~~0~~,~~0~~,.....,~~1 \\
         (-1)^{d+1},a_{1},a_{2},..,a_{d-1}

\end{array} \right) . \eqno(4)$$

\vspace{.5cm}

{\bf 3. }Let us consider trajectories of the system (1) 
with initial vector 
$X^{(0)}$ which has rational coordinates 

$$ X_{0} = (~~\frac{q_{1}}{p_{1}},~~\frac{q_{2}}{p_{2}},...,
\frac{q_{d}}{p_{d}}~~). \eqno(5)$$
It is easy to see, that all these trajectories 
are periodic orbits of the 
Anosov map (1), because matrix elements~~ $a_{i,j}$~~ are integer.
Indeed, if we shall consider the sublattice of unit torus $\Pi^{d}$ 
with rational coordinates of the form~~ $q/p$~~ where $~p~$ is 
the smallest common denominator

$$X=(~~\frac{q_{1}}{p},~~\frac{q_{2}}{p},...,
\frac{q_{d}}{p}~~), ~~~~~~0\leq q_{i} \leq p-1  \eqno(6)$$
then the multiplication,summation and $(mod)$ operations (1) will leave the 
trajectory on the same sublattice. The total number of vertices on this 
sublattice $Z_{p}^{d}$ is 

$$(total~number~of~verteces)= p^{d}, \eqno(7)$$
therefore the period $\tau_{p}$ of the trajectories on 
$Z^{d}_{p}=Z_{p}\otimes ...\otimes Z_{p}$, where 
$Z_{p}=\{0,1,...,p-1 \}$ 
is always less than $p^{d}$

$$\tau_{p} \leq p^{d}.$$
Thus the periodic trajectories of this system (1) with the initial vector 
(5) coincide with a subset of the 
points of rational sublattice $Z^{d}_{p}$ and 
our goal is to find conditions under which the period of the system will 
be as large as possible.

Let us show that on every given sublattice $Z^{d}_{p}$ Anosov map (1) 
reduces to 
($\it mod$ $p$) arithmetic. Indeed on sublattice $Z^{d}_{p}$ the Anosov
map $A$ (1) can be written as

$$\frac{ q^{(n+1)}_{i}}{p} =
\sum_{j}a_{i,j}~\frac{q^{(n)}_{i}}{p},~~~~~~~~~~(mod~~1) \eqno(8)$$ 
and is equivalent to ($mod~~p$) arithmetic on the lattice with 
integer coordinates $q_{i}$ which are in the interval $[0,p-1]$ 

$$q^{(n+1)}_{i} =\sum_{j}a_{i,j}~q^{(n)}_{i},~~~~~~~~~(mod~~p).\eqno(9)$$
Thus the images of the 
periodic trajectories on a unit torus $\Pi^{d}$ appear as trajectories
on the integer sublattice $Z^{d}_{p}$ and all operations can be 
understood ($mod~~p$). The most important thing is that now all 
operations become commutative.

\vspace{.5cm}

{\bf 4. }To estimate the period of the trajectories on rational 
sublattice it is essential to consider those sublattices 
for which $p$ is the prime number, we mean that $p_{1}=...=p_{d} =
p$. In that case the integer 
sublattice gains an additional structure and becomes the Galois 
field and all operations reduce to arithmetic ones on Galois field.
The benefit to work on Galois field is that four arithmetic 
operations are well defined on that sublattice \cite{apostol}.

In this way we can consider every coordinate $q_{i}$ , 
$i=1,...,d$ as belonging to Galois field 
$GF[p]=\{0,1,...,p-1\}$, where $p~is~prime~number$ and consider 
the sublattice as a direct product of Galois fields

$$Z^{d}_{p} = GF[p]\otimes...\otimes GF[p] .\eqno(10)$$
As we already mentioned, this reduction of the dynamical system (1)
to dynamical system for which the Galois field plays 
the role of the phase space makes all operations commutative in the 
sense that 

$$\{ A \{ A ~X \}\} = \{ A^{2}~X  \}, \eqno(11)$$
where $\{...\}$ means $mod$ operation.
The commutativity of the multiplication and $(mod)$ operation on the 
Galois sublattice means that the periodic trajectory 

$$\{ A \{A.........\{A~X\}...\}\} = X \eqno(12)$$
can be represented in the form 

$$ \{ A^{\tau_{p}}~X \} =X \eqno(13)$$
and the period of the trajectory $\tau_{p}$ can be understood as a 
degree of power on which the matrix $A$ reduces to identity ($mod~~p$)

$$A^{\tau_{p}} = 1~~~~~~~~~~~~~~(mod~~p) \eqno(14)$$
The period of the trajectory
on the Galois sublattice is equal therefore to the 
power $\tau_{p}$ in which the matrix $A$ reduces to identity 
on a given Galois field $GF[p]$. This period does not depend 
on initial vectors and the whole phase space $Z^{d}_{p}$
factories into trajectories with equal periods.
It is obvious that the same matrix $A$ will have different 
periods on different Galois fields and that this period 
depends on the given prime number $p$ and the dimension $d$ of 
matrices. 

\vspace{.5cm}

{\bf 5. }To demonstrate this fact let us consider few examples.
The matrix 
$$A =\left( \begin{array}{c}
         2,1 \\
         1,1
\end{array} \right) . \eqno(15)$$
has period equal to four on the Galois field with $p=3$
and to eight when $p=7$

$$ \tau_{3}(A)=4,~~~~~~~~\tau_{7}(A)=8.$$
The question which appears here, is how it is possible to estimate 
the period of the matrix $A$ without actual computation of the 
powers of the matrix $A$. 

We can find the answer to this question considering the eigenvalues 
of the matrix $A$. Indeed, as we will see, we can compute the 
periods using the eigenvalues of the matrix $A$.
Let us consider first the example of the eigenvalues of the cat map (15) 

$$\lambda_{+}= 
\frac{3+ \sqrt 5}{2},~~~~~~~\lambda_{-}= \frac{3-\sqrt 5}{2}$$
The question: what is the period of the given matrix $A$ 
on Galois field $GF[p]$ is equivalent now to the question :
in which power the eigenvalues are equal to identity on field $GF[p]$?

$$\lambda^{\tau_{p}} =1 ~~~~~~~~(mod~~p) \eqno(16)$$
As it is easy to see  

$$\lambda^{4}_{+}=1~~(mod~~3),~~~~\lambda^{8}_{+}=1~~(mod~~7)$$
which confirms the direct computation. 
The exceptional case 
when this method can not be applied directly is when the 
eigenvalues have 
degeneracy on a particular field $GF[p]$. This 
takes place for $p=5$,~~~indeed~~ $\lambda_{+} = 
\lambda_{-} = 4,~~(mod~~5)$.
Using Jordan normal form of the matrix one can see that $\tau_{5} = 10$.
Because this happens for very particular values of $p$ in 
the following we will consider  only the cases when 
eigenvalues are not degenerate.

\vspace{.5cm}

{\bf 6. }
Thus the actual value of the period $\tau_{p}$ naturally depends 
on the form of eigenvalues $\lambda$ and of the prime number $p$.
Here we can distinguish different cases:

\vspace{.5cm}

i). The eigenvalue $\lambda$ coincides with one of the elements of the 
Galois field $GF[p]$. In that case the period $\tau_{p}$ 
depends on whether eigenvalue coincides with the primitive element of the 
Galois field or not. All elements of the field $GF[p]$ can be 
constructed as powers of primitive element $g$ and $g^{p-1}=1$.
If one of the eigenvalues coincides with the primitive element 
of the Galois field , 

$$\lambda = g,~~~~~where~g~is~the~primitive~element~of~GF[p], \eqno(17)$$
then the period of the matrix is maximal and is equal to 
$\tau_{p} = p-1$
 
$$\lambda^{p-1} =1,~~~~~~~~~(mod~~p). \eqno(18)$$
Therefore to get the maximal  period in the case i)
one should have at least one of the eigenvalues equal to the 
primitive element of the field
$GF[p]$. If $\lambda$ does not coincide with the primitive element $g$, 
then the period is simply smaller. 
\vspace{.5cm}

ii). The eigenvalue does not coincide with any 
of the elements of the Galois field $GF[p]$. This takes place 
because the solutions of the characteristic polynomial of the 
matrix $A$ are not always in the field $GF[p]$. Galois field is 
arithmetically 
complete, but it is not algebraically complete, therefore one 
can have the situation when

$$\lambda ~~is~~not~~an~~element~~of~~GF[p]. \eqno(19)$$
This possibility can be illustrated by cat map (15), indeed  
$\sqrt 5$ is not an element of GF[3] or GF[7].

In that case one should ask, whether it is  an 
element of the quadratic extension $GF[\sqrt{p}]$. The 
quadratic extension of the Galois field consists of the numbers 
of the form $a+b\sqrt g$ where $a,b$ are the elements of field $GF[p]$, 
$g$ is the primitive element of $GF[p]$ and 
$\sqrt g$ is a square-free integer. 

Now if the eigenvalue is an element of the quadratic extension and 
coincides with it's primitive element $h$

$$\lambda=h,~~~~~where~~~h =
h_{1}+h_{2}\sqrt{ g}~~~~~~~is~the~primitive~element~of~GF[\sqrt{p}], 
\eqno(20)$$  
then the period is equal to $\tau_{p} = p^{2}-1$

$$\lambda^{p^{2}-1} =1,~~~~~~~~~(mod~~p) \eqno(21)$$
because the primitive element of the $GF[\sqrt{p}]$ has the period 
equal to $p^{2}-1$ \cite{apostol}. Again, if $\lambda$ does not 
coincide with the primitive element $h$, then the period is simply 
smaller, as it is in the case (15)  for $p=3,5$ where the period
is $p+1$.
\vspace{.5cm}

iii). In general the characteristic polynomial of the matrix $A$ is 
of order $d$ and the eigenvalues can belong to high extensions 
$GF[\sqrt[d]{p}]$ of the Galois field, the elements of which have
the form $a+bq+...+eq^{d-1}$~~where $a,b,...,e$ are the elements of 
$GF[p]$,~~ $g$ is the primitive element and $q^{d}=g$. The primitive 
element $h=h_{1}+h_{2}q+...+h_{d}q^{d-1}$ of $GF[\sqrt[d]{p}]$ has the 
period $\tau_{p}=p^{d}-1$

$$\lambda^{p^{d}-1}=1,~~~~~~~~(mod~~p). \eqno(22)$$ 
\vspace{.5cm}

{\it This analysis demonstrates an important fact that in order to have a 
large period on a 
sublattice $GF[p]$ one should have matrices $A$ which 
have the eigenvalues in high extensions $GF[\sqrt[d]{p}]$ of
the field and at the same time they should have integer entries.}

\vspace{1.0 cm}

{\bf 7. }In the previous sections we described the trajectories 
of the K system on the rational sublattice $Z^{d}_{p}$ and 
particularly 
on a Galois field, that is when p is the prime number.

In this section we will reverse the discussion and will try to 
construct the matrices A with the properties of K systems on 
a given Galois field with the maximal period. The question can be 
formulated in the following form: can we construct a matrix A 
with the properties of K system such that it has the largest period 
on a given Galois field $GF[p]$ ?

Let us first consider two-dimensional matrices of the form 

$$A_{2} =\left( \begin{array}{c}
         \alpha,~~~~~~ \alpha +1\\
         \alpha -1,~~\alpha 
\end{array} \right) ,~~~~(mod~~p) \eqno(23)$$
which have the following eigenvalues

$$\lambda_{+}=\alpha +\sqrt{ \alpha^{2}-1},~~~~\lambda_{-}=
\alpha -\sqrt{ \alpha^{2}-1}$$
To realize the first case i), when the eigenvalue belongs to the
field, we should have $\sqrt{\alpha^{2}-1}$ as an element of 
the field, that is 

$$\alpha^{2}-1 = k^{2},~~~~~~~~ k \neq 0. \eqno(24)$$
In this case the square root operation will belongs to the field.
To have the maximal period we should choose one of the 
eigenvalues to be the primitive element $g$ of the given field $GF[p]$

$$\lambda_{+}=g,~~~~\lambda_{-}=g^{-}, \eqno(25)$$
therefore

$$\alpha + k =g^{-},~~~~~~~\alpha -k =g \eqno(26)$$
and (24) is satisfied automatically. From (26) 

$$k=\frac{g^{-} - g}{2},~~~~~~~\alpha =
\frac{g^{-} + g}{2},~~~~~~~~g \neq g^{-},\eqno(27) $$
so that the matrix (23) is equal to

$$A_{2} =\left( \begin{array}{c}
         \frac{g^{-} + g}{2}~~~~~~~~~,\frac{g^{-} + g}{2} +1\\
                                                           \\      
          \frac{g^{-} + g}{2}-1,~~~~~\frac{g^{-} + g}{2}
\end{array} \right) . \eqno(28)$$
and has the period as large as the 
primitive element $g$, which is $p-1$

$$\tau_{p}(A_{2}) = p-1. \eqno(29)$$
The upper bound for the period of the cat map (15) 
found by Dyson and Falk \cite{dyson} 
on an arbitrary rational sublattice $Z_{q}$ is equal to 

$$\tau \leq 3q $$
and also linearly depends on $q$.
\vspace{.5cm}

{\bf 8. }These formulas allow to construct explicit examples of 
matrices with given period. The field $GF[7]$ has primitive 
element $g=3$ and $g^{-}=5$ therefore the matrix (28) has the form

$$A =\left( \begin{array}{c}
         4~~~5 \\
         3~~~4
\end{array} \right),~~~~~~~~~~ (mod~~7) \eqno(30)$$
with the corresponding period  $\tau_{7}=6$.

\vspace{1.0 cm}

{\bf 9. }
The next step in this construction is to enlarge the 
sublattice $GF[p]$ to sublattice $Z_{p^{n}}$, where p is 
the same prime number. Despite the fact that the sublattice $Z_{p^{n}}$ 
does not 
have field structure, nevertheless there exists an element 
$h$ with the period $(p-1)p^{n-1}$.
The important theorem \cite{apostol} states that~~ $h$~ coincides 
with one of the primitive elements 
of the original field $GF[p]$ which has the property

$$g^{p-1} \neq 1,~~~~~~(mod~~p^{2}). \eqno(31)$$
This primitive element $g$ is the same for $Z_{p^{n}}$ for any 
integer $n \geq 2$ and has the period 

$$g^{(p-1)p^{n-1}}=1~~~~~~~(mod~~p^{n}).$$
We have therefore the following result: the  matrices 
which we have constructed in the previous section on 
$GF[p]$, will have period on sublattice $Z_{p^{n}}$ equal  to 

$$\tau_{p^{n}}(A_{2})= (p-1)p^{n-1}~~~~~~~~(mod~~p^{n}). \eqno(32)$$
It simply means that with $(mod~~p^{n})$ operation we 
increase the period of the matrix $A_{2}$ from $p-1$ 
to $(p-1)p^{n-1}$. This allows to have large sublattices with  
small basic primes. 
\vspace{1.0 cm}

{\bf 10. }For the case $GF[7]$ the condition (31) is satisfied because 

$$3^{6} \neq 1,~~(mod~~49)$$
and the matrix (30) has the period 

$$\tau_{7^{n}}(A_{2}) = 6\cdot 7^{n-1}~for~any~ n>2~. $$
So our construction of the matrices  
with eigenvalues which are proportional to the primitive 
element of the field $GF[p]$ is completed. 
\vspace{1.0 cm}

{\bf 11. } As a basis for next constructions, let us 
consider a class of K system generators with very simple structure
\cite{nersesian}

$$A_{d} =\left( \begin{array}{c}
         0,~~1~,~~0~,.....,~~0 \\
         0,~~0~,~~1~,.....,~~0   \\
         .............   \\
         ............. \\
         0,~~0~~,~~0~~,.....,~~1 \\
         (-1)^{d+1},a_{1},a_{2},..,a_{d-1}
\end{array} \right) . \eqno(33)$$
In the last case it is easy to compute the characteristic 
polynomial of $A_{d}$

$$\lambda^{d}-a_{d-1}~\lambda^{d-1}-...-a_{1}~\lambda + (-1)^{d} = 
0 \eqno(34)$$
and therefore for it`s eigenvalues $\lambda_{1},...,\lambda_{d}$
we have 

$$\lambda_{1}\cdot \cdot \cdot \lambda_{d} =1$$
$$............$$
$$\lambda_{1} +...+ \lambda_{d}= a_{d-1} .\eqno(35)$$
These formulas allow to choose eigenvalues and then to 
construct matrix $A_{d}$ for K system generators.

For example if d=4 and $a_{1}=0,~a_{2}=3$,~and~$a_{3}=0$,
then

$$\lambda_{1} =\sqrt {\frac{3+\sqrt{5}}{2}},~\lambda_{2} 
= -\sqrt {\frac{3+\sqrt{5}}{2}},~ $$
$$\lambda_{3} =\sqrt {\frac{3-\sqrt{5}}{2}},~\lambda_{4} 
=-\sqrt {\frac{3-\sqrt{5}}{2}}, \eqno(36)$$
with an additional simplectic structure of $A_{4}$.
\vspace{1.0 cm}

{\bf 12. } Our goal is to get matrices of the form (33) with 
the maximal period. Let us first consider four-dimensional case

$$A_{4} =\left( \begin{array}{c}
         ~0,~~1~~,~~0~~,~~0 \\
         ~0,~~0~~,~~1~~,~~0   \\
         ~0,~~0~~,~~0~~,~~1 \\
         -1,~~0,~2\alpha,~0
\end{array} \right) . \eqno(37)$$
which has the characteristic polynomial 

$$\lambda^{4} -2\alpha \lambda^{2} +1 =0,\eqno(38)$$
and we will choose again $2\alpha=g+g^{-}$, then the roots are:

$$\lambda_{1,2}=\pm \sqrt{g} ,~~~\lambda_{3,4}=\pm \sqrt{g^{-}}.
\eqno(39)$$
Because $g$ is square-free primitive element of the 
$GF[p]$ the period of this matrix is  

$$\tau_{p}=2(p-1). \eqno(40)$$
Increasing the dimension of the matrix $A$ with the same 
simplectic structure

$$A_{2d} =\left( \begin{array}{c}
         0,~~1~,~~0~,.....,~~0 \\
         0,~~0~,~~1~,.....,~~0   \\
         .............   \\
         ............. \\
         0,~~0~~,~~0~~,.......,~~1 \\
         -1,~~0~~,~~0~~,...2\alpha,..,~0
\end{array} \right) . \eqno(41)$$
we will get the characteristic polynomial

$$\lambda^{2d} -2\alpha \lambda^{d} +1 =0,~~~~2\alpha=g+g^{-}\eqno(42)$$
with $2d$ different roots 

$$\lambda^{d}_{+}=\alpha+\sqrt{\alpha^{2}-1},~~~~~\lambda^{d}_{-}=
\alpha -\sqrt{\alpha^{2}-1}\eqno(43)$$
and the period is equal to 

$$\tau_{p}(A_{2d}) = d(p-1). \eqno(44)$$
The same matrices $A_{2d}$ on $Z_{p^{n}}$ sublattice will have 
the period

$$\tau_{p^{n}}(A_{2d}) = d(p-1)p^{n-1}. \eqno(45)$$
\vspace{1.0 cm}

{\bf 13. } The example on $GF[7]$ where $g+g^{-}=8=1=-6$ is 

$$A_{2d} =\left( \begin{array}{c}
         0,~~1~,~~0~,.....,~~0 \\
         0,~~0~,~~1~,.....,~~0   \\
         .............   \\
         ............. \\
         0,~~0~~,~~0~~,.......,~~1 \\
         -1,~~0~~,~~0~~,..-6,..,~0
\end{array} \right),~~~~~~~~~~~~(mod~~7)  \eqno(46)$$
with period $\tau_{7}=d\cdot 6$ and with $(mod~~7^{n})$ 
we have $\tau_{7^{n}}=d\cdot 6\cdot 7^{n-1}$.
\vspace{1.0 cm}

{\bf 14. }The next step is to construct the matrices which have the 
eigenvalues in quadratic extension  $GF[\sqrt{p}]$, 
that is we are going to consider the case ii). 
If $h$ is the primitive element of the $GF[\sqrt{p}]$, that is 

$$h=h_{1}+h_{2}\sqrt{g},~~~~~~~h\cdot 
h^{\star}=g,~~~~ h+ h^{\star}=2h_{1},\eqno(47)$$
then the matrix which has the eigenvalues in $GF[\sqrt{p}]$ can be 
constructed in the same form as (33) 

$$A_{3}= \left( \begin{array}{c}
   0,~~~~~~~~~~~~1,~~~~~~~~~~~~~~~0 \\    
   0,~~~~~~~~~~~~0,~~~~~~~~~~~~~~~1\\ 
   -1,~~~~~~2h_{1}g^{-}-g,~~~~2h_{1}-g^{-}     
\end{array} \right),~~~~(mod~~p) \eqno(48)$$
because the characteristic equation is 

$$(\lambda +g^{-})(\lambda -h)(\lambda-h^{\star})=$$
$$\lambda^{3}-(2h_{1}-g^{-})\lambda^{2}-
(2h_{1}g^{-}-g)\lambda +1=0~~~~(mod~~p) \eqno(49)$$
and has the root $h$ which coincides with the primitive element 
of $GF[\sqrt{p}]$. 
This matrix has integer elements by construction and the period  

$$\tau_{p}(A_{3})=p^{2}-1. \eqno(50)$$
Therefore the period quadratically increases in comparison with previous 
construction. 
The same matrix with $(mod~~p^{n})$ operation will give 

$$\tau_{p^{n}}(A_{3})=(p^{2}-1)p^{n-1}. \eqno(51)$$
\vspace{1.0 cm}

{\bf 15. } The example on $GF[7]$ where $h=1+2\cdot \sqrt{3}$ will be 

$$A_{3}= \left( \begin{array}{c}
   0,~~~~~~1,~~~~~~0 \\    
   0,~~~~~~0,~~~~~~1\\ 
   -1,~~~~~~0,~~~~~~4     
\end{array} \right),~~~~(mod~~7) \eqno(52)$$
with $\tau_{7}=48$ and for $(mod~~7^{n})$ the period is 
$\tau_{7^{n}}=48\cdot 7^{n-1}$.
\vspace{1.0 cm}

{\bf 16. }To construct the matrix generator with eigenvalues 
in high fields $GF[\sqrt[d]{p}]$ it is easier to use primitive 
polynomial of degree $d$ over $GF[\sqrt[d]{p}]$ the root of 
which coincides with the primitive element $GF[\sqrt[d]{p}]$.
The primitive polynomial has the 
form \cite{apostol,lidl,bastida}

$$\lambda^{d}+\beta_{1} \lambda^{d-1}+\beta_{2}\lambda^{d-2}+...
+\beta_{d} =0\eqno(53)$$
with coefficients $\beta_{1},\beta_{2},...,\beta_{d}$ over $GF[p]$. 
The only problem is that this 
polynomial does not correspond to a matrix with unit determinant (2). 
But the last term $\beta_{d}$ always 
can be represented as a power of the primitive element 
$g$ of $GF[p]$~~~$\beta_{d}=g^{k}$, therefore if we multiply 
the primitive polynomial (53) by $\lambda+g^{-k}$ we will get the 
polynomial which corresponds to a matrix with unit determinant

$$(\lambda +g^{-k})(\lambda^{d}+\beta_{1} \lambda^{d-1}+
\beta_{2}\lambda^{d-2}+...+\beta_{d})=$$
$$\lambda^{d+1}+(\beta_{1}+g^{-k})\lambda^{d}+
(\beta_{2}+\beta_{1}g^{-k})\lambda^{d-1}+...
+1=0.\eqno(54)$$
To this polynomial corresponds the matrix
generator of the form (33) 

$$A_{d+1} =\left( \begin{array}{c}
         0,~~1~,~~0~,...................................,~~0 \\
         0,~~0~,~~1~,...................................,~~0   \\
         .............   \\
         ............. \\
         0,~~~~~~...~~~~~~~~~~0~~~~~~~~~,~~~~~~~~~1~~~~~ \\
         -1,.....,-(\beta_{2}+\beta_{1}g^{-k}),~~~~-(\beta_{1}+g^{-k})
\end{array} \right)~~~~~~(mod~~p) . \eqno(55)$$
with period 

$$\tau_{p}(A_{d+1})=p^{d}-1 \eqno(56)$$
and on $Z_{p^{n}}$ 

$$\tau_{p^{n}}(A_{d+1}) =(p^{d}-1)p^{n-1}\eqno(57)$$
{\it This is our main result with the largest period of order $p^{d+n}$.}
\vspace{1.0 cm}

{\bf 17. }The example of the primitive polynomial on $GF[7]$ 
with $d=10$ is \cite{lidl}~~~ $\lambda^{10}+\lambda^{9}+
\lambda^{8}+3=0 $~~~~and (54) has the form $\lambda^{11}-
\lambda^{10}-\lambda^{9}-
2\lambda^{8}-4\lambda+1=0$~~~therefore the matrix is 

$$A_{11} =\left( \begin{array}{c}
         0,~~1~,~~0~,.....,~~0 \\
         0,~~0~,~~1~,.....,~~0   \\
         .............   \\
         ............. \\
         0,0,0,...,0,0,0,1 \\
         -1,4,0,...,0,2,1,1
\end{array} \right)~~~~~~(mod~~7) \eqno(58)$$
with period $7^{10}-1$ and $(7^{10}-1)7^{n-1}$ on $Z_{p^{n}}$.  

It is also useful to have the list of 
primitive polynomials on $GF[2]$  \cite{lidl}. Tables with larger ranges
of $d$ are available for $GF[2]$. In particular \cite{watson}
contain tables for $d<101$, in \cite{stahnke} for $d<169$ and 
in \cite{brillhart} for $d<1001$ with the corresponding period
$2^{1000}-1$

$$d=24~~~~~~~~~~~~\lambda^{24}+\lambda^{23}+\lambda^{22}+\lambda^{17}+1=0$$
$$d=81~~~~~~~~~~~~~~~~~~~~~~~~~~~~~~\lambda^{81}+\lambda^{4}+1=0$$
$$d=97~~~~~~~~~~~~~~~~~~~~~~~~~~~~~~\lambda^{97}+\lambda^{6}+1=0$$
$$d=127~~~~~~~~~~~~~~~~~~~~~~~~~~~~~~\lambda^{127}+\lambda+1=0$$
$$d=159~~~~~~~~~~~~~~~~~~~~~~~~~~~~~~\lambda^{159}+\lambda^{31}+1=0$$
$$d=165~~~~~~~~~~~~\lambda^{165}+\lambda^{31}+\lambda^{30}+\lambda +1=0$$
$$d=167~~~~~~~~~~~~~~~~~~~~~~~~~~~~~~\lambda^{167}+\lambda^{6}+1=0.$$
In this case one can directly construct the matrices of 
the form (33) because the free term is equal to one. 
For the last polynomial we have 

$$A_{167} =\left( \begin{array}{c}
         0,~~1~,~~0~,.....,~~0 \\
         0,~~0~,~~1~,.....,~~0   \\
         .............   \\
         ............. \\
         0,.....0,0,0,0,0,1 \\
         1,.....1,0,0,0,0,1
\end{array} \right)~~~~~~(mod~~2) \eqno(59 )$$
with period $\tau_{2}(A_{29})=2^{167}-1$. Direct check 
of the eigenvalues, shows that eigenvalues are not on a unit 
circle, therefore the K conditions (2,3) are satisfied. 
We have checked that for all primitive polynomials on hand 
this conditions are satisfied, so one can use any of them.
\vspace{1.0 cm}

{\it Acknowledgments} G.K.S. acknowledge G.Pavlos and M.Bountouridis 
for kind hospitality at the Interdisciplinary Center for Complexity of the 
Research Foundation of Thrace.

\vfill
\end{document}